\documentclass[12pt,preprint]{aastex}

\newcommand{\ud}{{\rm d}}
\newcommand{\Teff}{$T_{\rm eff}$}
\newcommand{\Vturb}{$V_{\rm turb}$}
\newcommand{\logg}{$\log g$}

\shorttitle{Theoretical limb darkening for Cepheids.}
\shortauthors{M. Marengo et al.}

\begin{document}


\title{Theoretical Limb Darkening for Pulsating Cepheids.}

\author{Massimo Marengo, Dimitar D. Sasselov, Margarita Karovska,
Costas Papaliolios}
\affil{Harvard-Smithsonian Center for Astrophysics, 60 Garden St.,
Cambridge, MA 02138}
\email{mmarengo,dsasselov,mkarovska,cpapaliolios@cfa.harvard.edu}

\and
\author{J.T. Armstrong}
\affil{Remote Sensing Division, Naval Research Laboratory, Code 7210,
Washington, DC 20375}
\email{tarmstr@gemini.usno.navy.mil}

\begin{abstract}

Ground-based interferometry has finally reached a stage in which
accurate determination of Cepheid diameters using the Baade-Wesselink
method is feasible. Determining these diameters is the base for
calibrating the period-luminosity relation for Classical Cepheids, and
thus the Extragalactic Distance Scale, but requires accurate limb
darkened models. 
This work presents a new method to compute time and wavelength
dependent center-to-limb brightness distributions for Classical
Cepheids. Our model atmospheres are based on second-order accurate 1-D
hydrodynamic calculations, performed in spherical geometry. The
brightness intensity distributions, and the resulting limb darkening,
are computed through the dynamic atmospheres by using a full set of
atomic and molecular opacities. 
Our results confirm important differences with respect to equivalent
hydrostatic models. The amount of limb darkening, and the shape
of the limb profiles, show a strong dependence on the pulsational
phase of the Cepheid, which cannot be reproduced by static models. 
Non-linear effects in our hydrodynamic equations add a new level of
complexity in the wavelength dependence of our limb profiles, which
are affected by the presence of shock-waves traveling through the
atmosphere. These effects, already detectable by present-day
interferometers, should be taken into consideration when deriving
limb darkened diameters for nearby Cepheids with the accuracy required
to measure their radial pulsations.

\end{abstract}




\section{Introduction}\label{sec-intro}

Classical Cepheids play a central role in setting the Extragalactic
Distance Scale. In fact, Cepheid high intrinsic luminosity, and the
existence of a well studied relation between their pulsation period
and luminosity (first discovered by \citealt{leavitt1906}), makes them
crucial primary distance indicators. These properties have been
exploited extensively, from early surveys to the recently completed
HST Key Project to measure the Hubble constant \citep{freedman2001}.

The reliability of Cepheids as distance indicators depends on the
accurate calibration of the period-luminosity relation
zero-point. This is difficult to accomplish, due to the low 
density distribution of Cepheids in our galaxy.
Direct determination of the trigonometric
parallaxes of the nearest Cepheids has been attempted using the database
of the HIPPARCOS mission (see e.g. \citealt{feast1997}), but large
errors do not allow an unambiguous interpretation of the results
\citep{madore1998}. An accurate measurement of the parallaxes of even
the closest galactic Cepheids does not appear feasible, even with the
space based surveys proposed for the near future \citep{horner2000}.
Most promising, among the techniques which can provide an independent
calibration of the period luminosity zero-point, is the
Baade-Wesselink (BW) method \citep{baade1926, wesselink1946}. 

Recent progress in ground-based
interferometry finally allows direct detection of pulsations in nearby
Cepheids \citep{lane2000}. The measurement of such radial angular
variations, combined with spectroscopic observations yielding the
radial velocity of the pulsating photosphere, can in principle allow a
precise determination of nearby Cepheid distances \citep[][hereafter
SK94]{sasselov1994b}. As described in SK94, the feasibility of such a
measurement is conditioned on the availability of accurate
hydrodynamic models for the Cepheid atmospheres. These are required
(1) to derive the projection factor necessary to convert the photospheric
line velocities into radial motion, and (2) to compute reliable
time-dependent limb darkening profiles. The first of these
requirements is described in SK94 and addressed in detail by
\citet{sabbey1995}. In this paper we describe a new approach to solve
the second of these requirements. 

Currently, limb darkening tables specifically computed for Cepheid
atmospheres are not available, so coefficients derived for static
atmospheres of non-variable yellow supergiants are generally used
\citep{parsons1971, manduca1977, manduca1979, kurucz1993,
claret1995}. This follows the assumption that a Cepheid atmosphere at
a given phase can be approximated with a static plane-parallel
atmosphere having the same \Teff{} and \logg, as estimated by fitting
the depth ratio of weak photospheric lines. This is generally true
as long as the observational constraints do not demand a level of
precision that is affected by the time-dependent hydrodynamics and by
the spherical geometry of the real pulsating star. 

The assumptions of
hydrostatic equilibrium and plane parallel geometry lead to an
underestimation of the temperature gradients in yellow supergiants,
and thus fail to reproduce the correct IR and UV fluxes
\citep{fieldus1990}. They also cannot reproduce the relative
separation between the continuum photosphere and the line-forming
region, and thus underestimate the difference in the radial-velocity
amplitudes \citep{karp1975, sasselov1992}. All these and other
limitations of static plane-parallel atmospheres seriously affect the
precision of Cepheid distance determination with the BW
method. Current demands on a Cepheid distance scale accuracy of 10\%
and better require us to abandon the approximation of hydrostatic
equilibrium and plane-parallel geometry (see SK94).

In \citet{sasselov1994} new time-dependent, non-LTE dynamic models of
Classical Cepheids were presented, and \citet{sabbey1995} used them to
study spectral line formation. In our paper, updated models are
used to further explore the consequences of the dynamic treatment of
Cepheids pulsation in the resulting spectral brightness distribution
and limb profiles. This last step is complex,
since a consistent model, able to treat at the same time
non-LTE radiative transfer and spherical geometry with all the
necessary opacities and line sources in a time-dependent
hydrodynamic framework, is still not feasible. Section~\ref{sec-descr}
illustrates our solution to this problem. 

In section~\ref{sec-disc} the typical results for a
classical Cepheid are analyzed, and the differences between our
computed limb darkened profile and the one from an equivalent static
atmosphere are discussed. The wavelength dependence of our limb
profiles is also analyzed. In section~\ref{sec-concl} we finally
discuss the applications of our models, and their implication on the
analysis of presently available and future interferometric measurements.


\section{Spectral Intensities for Dynamic Atmospheres}\label{sec-descr}

The computation of a realistic spectral intensity distribution for a
stellar model requires a complete set of opacities for the atmospheric
atomic species. If \Teff $\la 7000$~K (which is generally true for
classical Cepheids),  molecules should also be included. Ideally, this
requires solving the full radiative transfer problem and the hydrodynamic
equations in a self consistent way. However, the needed computing
power is excessive, and therefore the hydrodynamic structure of the
atmosphere is computed by considering only a limited set of elements
(see discussion in section~\ref{ssec-dmods}). The full set of
elements and their opacities should somehow be reintroduced when
deriving the emergent spectra.

A possible approach consists in using the
Rosseland opacities derived in hydrostatic equilibrium. The
hydrostatic equivalent of the dynamic model is chosen to match
as close as possible the dynamic atmosphere, as it changes during the
Cepheid pulsation. For each snapshot of the hydrodynamic simulation,
one has to search for a static LTE model producing the same total
flux of the star, and having a similar atmospheric structure. 
A ``hybrid'' model is thus artificially constructed,
having the depth-dependent opacities of the static atmosphere, and the
temperature, electron density and pressure of the dynamic model
atmosphere. 

The main limitation of this procedure is that the opacities are still
computed  in LTE conditions. However, the intensity distributions thus
derived are much improved, since the thermodynamical state is set by the
hydrodynamical calculations. The accuracy and caveats of this
technique are analyzed and discussed in the following sections.


\subsection{Dynamic Models}\label{ssec-dmods}

Hydrodynamic simulations show fundamental departures in the structure
of pulsating atmospheres from hydrostatic equilibrium and LTE 
\citep{willson1985, cuntz1987}.  Our hydrodynamical calculations only
include model ions of H, \ion{He}{1} and \ion{}{2}, \ion{Ca}{2} and
\ion{Mg}{2} \citep[see][for details]{sasselov1994}. Additional
non-LTE calculations are then performed with model ions of \ion{C}{1}
and \ion{}{2}, \ion{O}{1}, \ion{Mg}{1}, \ion{Ca}{1} and \ion{Fe}{1}. 
These additional species do not provide any feedback for the
radiative hydrodynamic simulations and are thus considered only for the
starting models, and after the dynamic calculations have converged. For
galactic Cepheids, Solar composition is assumed for all elements.

The pulsation is introduced by perturbing the atmosphere
with a ``piston'' having suitable period, amplitude and shape. These
parameters are specific for each Cepheid under study, and are derived
from observational constraints. The period of the piston is the period
of the Cepheid. The shape of the piston is taken from linear
non-adiabatic model computations of \citet{buchler1990}. The piston
amplitude is the main free parameter, and is derived by matching the
radial velocity curve, the phase lags, the estimated extension of the
atmosphere, the observed color and light variations and other basic
observable features. When the oscillations introduced by the piston
stabilize, and no transients remain, a set of time-dependent pulsating
photospheric models are computed for the entire pulsational cycle. 
The details of this procedure are explained in \citet{sasselov1994}. 

Given a good model for the Cepheid piston, it is
thus possible to fit the main stellar observables with a set of
dynamic models, each one representing a snapshot of the atmosphere at
any pulsational phase. We have applied this procedure for $\zeta$
Gem, which is a typical classical Cepheid. 

We note that in \citet{sasselov1994} a further step is performed,
which consists of building a semi-empirical chromosphere to derive the
phase-dependent profile of individual chromospheric lines. Since we
are mostly interested in deriving low spectral resolution intensities
and visibilities, the computed chromosphere is ignored in this work,
due to its negligible contribution to the overall limb darkening
profiles.


\subsection{Dynamic Models and Stellar Observables}\label{ssec-fourier} 

The Cepheid
pulsation leads to the extension and contraction of the atmosphere, 
responsible for the changes of the stellar observables (\Teff,
photospheric lines velocity, radius, luminosity, etc). These
quantities, which change with the phase of a pulsating star,
parameterize the computation of its phase-dependent spectrum. Let us
define a consistent reference framework for the phases of a Cepheid
cycle. 

Traditionally, the zero-point phase of variable stars is set to
coincide with maximum luminosity. This choice, however, is not
convenient when dealing with time-dependent hydrodynamic models, since
the emergent flux in the visible is a derived quantity, not always
available in the numerical computation. A better approach is to use
hydrodynamic quantities directly related to observables. The common
choice is the pulsational velocity of the Cepheid photosphere, which
can be observed by measuring the Doppler shift of suitable photospheric
lines. 

Figure~\ref{fig-curves} (left) shows the radial velocity and
lightcurve of $\zeta$~Gem. These were derived by tables published by
\citet{bersier1994a, bersier1994b}, where these parameters are
measured for a large set of Cepheids, and expressed in terms of their
Fourier coefficients.  The 
conversion of radial velocities into pulsational motion requires us to
perform the following geometric transformation: 

\begin{equation}\label{eq-vp}
v_p \simeq - p \left( v_r - \gamma \right)
\end{equation}

\noindent
where $\gamma$ is the systemic velocity taking into account the radial
velocity of the star relative to the solar system, the minus sign
inverts Doppler shifts into velocities relative to the Cepheid
atmosphere, and $p$ is the projection factor. The pulsational velocity
of $\zeta$~Gem derived using Equation~\ref{eq-vp} is plotted
in Figure~\ref{fig-curves} (right-top panel), using a constant 
projection factor of 1.36 \citep{burki1982}. The systemic velocity
$\gamma = 5.91$~km/s was determined by requiring conservation of
radius after the completion of one pulsational cycle, and is
consistent with the value of $6.5(\pm 1.0)$~km/s derived by
\citet{wilson1953} for the radial velocity of the star.

The zero phase in the velocity curve, and in the subsequent
discussion, is set at the time in which $v_p$ passes from negative to
positive, e.g. when the star begins to expand after reaching the
minimum radius. This point is marked in Figure~\ref{fig-curves} with a
solid vertical line. In the case of $\zeta$~Gem, the maximum
luminosity occurs at phase $\phi \simeq 0.28$, which is 3 days after
the zero-point phase (dashed line), given its 10.15~days period.
Minimum luminosity is at $\phi \simeq 0.76$ (dotted
line), while maximum radius is at $\phi \simeq 0.59$ (dot-dashed
line). 

The radius R is obtained by integrating the velocity $v_p$ over time: 

\begin{equation}\label{eq-radius}
R(t) = R_0 + \int_{t_0}^t v_p(t') \ud t'
\end{equation}

\noindent
where $R_0$ is the mean radius. An example is shown in
Figure~\ref{fig-curves} (bottom-right) for $\zeta$~Gem, which has $R_0
\simeq 69$~R$_\odot$ \citep{krockenberger1996}.

Note, however, that the above expression is strictly correct only if
$v_p$ is associated with an individual mass element in the stellar
atmosphere. This is not the case for the photospheric line
velocities (as the lines are produced by different photospheric layers
during the stellar pulsation), nor for the velocity field of the
hydrodynamic model (as the radial grid is not comoving with the
atmosphere). This inconsistency, which translates in the
non-conservation of the path integral in equation~\ref{eq-radius} over
one pulsational period, should be taken into account in order to have
a consistent reference frame for the observed and model velocities. 

The condition of constant average radius for the pulsating atmosphere
can however be achieved by adding an extra factor to the $\gamma$
constant in Equation~\ref{eq-vp}. This was done implicitly
when deriving the systemic velocity for $\zeta$~Gem.
We did the same with the model pulsational velocity,
deriving a correction factor of $-0.475$~km/s. With this condition,
the model and observed $v_p$ are directly comparable, allowing to set
the zero point phase in the sequence of models. 

The pulsational velocity of the hydrodynamic model was defined as the
average velocity of the photospheric layers, weighted over the region
where the line formation occurs. This quantity, shown in
Figure~\ref{fig-curvesk}, is very similar to the observational
$v_p$. Note the perturbation starting at $\phi \simeq 0.9$, which is
associated with a shock-wave crossing the photosphere. 

The bottom panel of Figure~\ref{fig-curvesk} shows the
effective temperature of $\zeta$~Gem as a function of phase, derived
by fitting the depth ratio of weak metallic lines as \ion{Fe}{1},
\ion{Fe}{2}, \ion{Si}{1} and a few other elements insensitive to the
changes in metallicities \citep{krockenberger1997}. The effective
temperature here defined is the \Teff{} of the plane-parallel LTE
atmosphere used to fit the observed line ratios. We will maintain this
convention through the following discussion, since this quantity
allows direct comparison between a model parameter and the
observations. 

The effective temperature plotted in Figure~\ref{fig-curvesk} is
related to the luminosity curve of the Cepheid, and is thus an
important observable that we use to constrain our model intensity
spectra, as explained in the following section.


\subsection{Rosseland Opacities for Dynamic Atmospheres}\label{ssec-opc}

The largest collection of stellar atomic and molecular opacities is
available as part of the ATLAS9$-$12 numeric code \citep{kurucz1970,
kurucz1979, kurucz1993b}, which incorporates no less than 58 million
lines for different stellar metallicities and values of microturbulent
velocity.

The Kurucz code initially computes static LTE model atmospheres in
plane-parallel geometry and radiative equilibrium. 
The input parameters which define each model are the effective
temperature \Teff, the gravity \logg, the convective mixing
length $L/H$, the turbulent velocity \Vturb{} and the
metallicity. The structure of a model atmosphere is evaluated on a
radial grid in the mass coordinate $\ud M = \rho \, \ud r$. For each
depth the model computes the local temperature, total pressure,
electron density, Rosseland opacity, radiation pressure and turbulent
velocity (which is not allowed to change along the radial grid). The
SYNTHE program, included in the ATLAS package, then allows the
computation of intensity spectra $I_\nu(\lambda,\mu)$, from which limb
darkened profiles are derived. 

In this work we adopted the ATLAS9 version of the Kurucz code, which
incorporates a full treatment for line-blanketing; we do not expect
that adopting ATLAS12 model atmospheres with opacity sampling would
have produced significant differences in the atmospheric structure
and resulting intensity profiles, at least in our wavelength range of
interest (Kurucz, private communication).

Figure~\ref{fig-A1} shows the main differences between the
snapshot of a dynamic atmosphere and a family of ATLAS9 models having
\logg{} from 0.5 to 2.0 and \Teff{} $\simeq 5600$~K. The effective
temperature for the static model has been chosen to match the
observational estimate for the Cepheid \Teff{} at the same phase. Contrary
to the hydrostatic case, the dynamic model shows a temperature
inversion for $\log \ud M \la 1.5$. This is due to the chromosphere
attached to the Cepheid atmosphere, which is not
reproduced by the static radiative equilibrium models. 
The deeper layers of the atmosphere,
on the contrary, can be better approximated by hydrostatic
equilibrium, with an appropriate choice in the \logg{} parameter. 

The agreement between the temperature distribution in the static and
dynamic atmospheres, below the temperature inversion point, depends 
on how significant are the departures from hydrostatic and radiative
equilibrium. A minor role is also played by departures from
plane-parallel geometry. As shown in \citet{sasselov1994}, however,
spherical geometry becomes an important factor only for long period
Cepheids, and can be neglected for our model of $\zeta$~Gem and the
other short period Cepheids that can be observed with the
available interferometers.

Consider phase $\phi \simeq 0.40$ (figure~\ref{fig-A1}, top panel),
which is close to average radius, when the star is freely
expanding. This is
well after the effects of the shock-wave (occurring at $\phi
\simeq 0.9$) dissipated. The dynamic and static thermal structures are 
very similar, at least where the convective overshooting
incorporated in ATLAS9 (absent in the dynamic models) is not
important. The overshooting only affects the deepest layers in the
stellar atmosphere, and is responsible for the lower temperature
gradient in the hydrostatic temperature profile.

At other phases (e.g. for $\phi \simeq 0.94$), the hydrodynamic
effects (the shock-wave in particular) are stronger, causing a larger
departure from hydrostatic equilibrium. (Figure~\ref{fig-A1}, bottom).
The spread in the temperature distribution in this case can be large,
and a single hydrostatic model which is a good approximation of the
hydrodynamic atmosphere cannot be identified.

In a typical Cepheid atmosphere, the contribution function for the
photospheric lines is negligible both in the chromosphere, and in the
deep layers subjected to convective overshooting, as shown in
\citet{sabbey1995}. In fact, the region  
where the hydrodynamic thermal structure is best approximated by an
equivalent hydrostatic atmosphere, is precisely where the photospheric
contribution functions are at maximum. Our assumption is that the
opacities computed in hydrostatic equilibrium in this region can be
used to compute emergent spectral intensities with the dynamic model.

On these basis, we fit our dynamic model for $\zeta$~Gem with a grid
of ATLAS atmospheres having \Vturb~$= 4$~km/s and \logg{} from 0.5 to
2.0 in steps of 0.1. Since the effective temperature for our
hydrodynamic models is not well defined, we have set the ATLAS
fitting grid to the \Teff{} value determined by
\citet{krockenberger1997} for each pulsational phase.
The fit was done by minimizing the normalized $\chi^2$, computed on the
``fitting region'' in which the contribution functions are at maximum,
between each dynamic model and all static atmospheres in the ATLAS
grid. This technique allows to find the best match between the
dynamic models and the static atmospheres, even in the cases in which
the dynamic model temperature structure cuts across several static
models (as shown in Figure~\ref{fig-A1} for $\phi = 0.94$). 
The result is a family of static atmospheres, each having a
Rosseland opacity computed with the full set of relevant
atomic species, to be used instead of the unknown opacities for the
dynamic models. The error associated with this procedure
can be visualized as the separation between the opacities of
two consecutive models in the fitting grid, which is $\la \pm 0.3$ in
the logarithmic scale (see figure~\ref{fig-A2}).

Note that, once the best fit static model is determined for each phase,
it is the whole atmospheric thermodynamic status that is constrained,
e.g. not only the temperature distribution $T$, but also the
electron density $x_{\rm ne}$ and the pressure $P$. Differences in the
$x_{\rm ne}$ and $P$ distributions between static and dynamic models
are to be expected, given the different equation of state used in the
two cases. These inconsistencies reflect in the accuracy of the
Rosseland opacity adopted for the dynamic atmosphere, which is
computed with the pressure and electron density relative to the
hydrostatic case.


\subsection{Intensities and Limb Profiles}\label{ssec-int}

The spectral intensity distribution for each phase of our dynamic
atmospheres are finally computed by adapting SYNTHE/ATLAS 
to process our ``hybrid'' models which combine the hydrodynamic
atmospheres with hydrostatic Rosseland opacities.

Each ``hybrid'' model is characterized by the following parameters:
(1) \Teff{} is the effective temperature of the best fit ATLAS model,
which was set to the value derived by \citet{krockenberger1997}; (2)
\logg{} determined by the best fit of the dynamic model with the
static grid; (3) \Vturb, $L/H$, metallicity and Rosseland opacities
taken from the best fit hydrostatic model; (4) $T$, $x_{\rm ne}$ and
$P$ obtained by splicing the hydrodynamic model (below the temperature
inversion point) with the static model (above $T$ inversion,
e.g. effectively removing the chromosphere). The removal of the
chromosphere is necessary, because SYNTHE is in essence a radiative
equilibrium code and cannot
treat temperature inversions and discontinuities in the radial
grid. This step is justified, as noted before, by the 
consideration that the only part of the atmosphere relevant
in the photospheric emission, is where the
contribution function is maximum. This is the same region used in
section~\ref{ssec-opc} to perform the fitting between static and
dynamic atmospheres. Eventual discontinuities around the temperature
inversion point in the two segments of the $T$, $x_{\rm ne}$ and $P$
profiles have been eliminated by rescaling the static part with a
suitable normalization factor.

An example of final temperature, electron density and pressure radial
profiles of our $\zeta$~Gem model is shown in
figure~\ref{fig-splice} for two representative phases. At $\phi \simeq
0.40$, the 
$T$ and $x_{\rm ne}$ dynamic model distributions closely match their
hydrostatic approximation, at least in the region used for the fit
(marked in the figure by the vertical dotted lines). The pressure of
the dynamic atmosphere, on the contrary, is lower than the static
equivalent, reflecting the change in
the equation of state. At phase 0.94, the extra
energy dissipated by the traveling shock wave is responsible for an
increase in the dynamic model electron density and pressure, which
becomes higher than in the hydrostatic atmosphere with the same
temperature profile. The increased electron density
of the ``hybrid'' model, reflecting the structure of the hydrodynamic
atmosphere, is in part responsible for the different limb profile of
the Cepheid at this phase. 

The spectral intensity distributions (and the associated limb
profiles) are then computed (together with a consistent set of
molecular opacities) by running SYNTHE on the ``hybrid'' models. The
results are discussed in the following section.


\section{Discussion}\label{sec-disc}

Limb profiles for our model Cepheid are obtained by plotting the
intensity $I_\nu(\lambda,\mu)$ as a function of the projected radial
coordinate $\sin \alpha$ (where $\alpha$ is the angle between the line
of sight and the direction of the emergent flux, and $\mu = \cos
\alpha$). Given the detailed treatment of stellar opacities in our
computation, the profiles thus obtained can be trusted 
analyzing the wavelength dependence of limb darkening. Also, the
explicit time dependence of our dynamic models allows us to follow the
variations in the limb profile as the Cepheid star pulsates.

\subsection{Differences between static and dynamic
models}\label{ssec-limbds} 

Figure~\ref{fig-limb} compares static and dynamic models at three
representative phases. At $\phi \simeq 0.4$ the limb darkening
profiles computed with the hydrodynamic models are very similar to the
hydrostatic equivalent. This is not surprising, since at this phase
the atmosphere is quasi-static, having long dissipated the transient
hydrodynamic effects caused by the passing shock-wave. Even though the
pressure is very different in the static and dynamic case, the
spectral emission (and thus the limb darkening) is similar, confirming
that $T$ and $x_{\rm ne}$ are the main parameters constraining the
photospheric emission. This is an important validation of the
procedure described in section~\ref{sec-descr}.

Dramatic evidence that the limb darkening is substantially affected
by the hydrodynamic effects is given in the two lower panels of
Figure~\ref{fig-limb}. The two panels compare the static and dynamic
limb profiles at $\phi \simeq 0.92$ (shock-wave crossing the
photosphere) and at $\phi \simeq 1$ (minimum radius). In the first
case the newborn shock-wave reduces $\nabla T$ with respect to the
hydrostatic atmosphere, and the limb darkening is lower. At minimum
radius the effects of the shock-wave have dissipated, and the
compression of the atmosphere gives rise to a
larger temperature gradient. This is responsible for a larger limb
darkening. In both cases the changes in the amount of limb darkening
with respect to the static model is mostly visible at shorter
wavelengths, where the opacity is lower and the limb darkening
larger.

\subsection{Changes of limb darkening with
phase and wavelength}\label{ssec-limbphase} 

Figure~\ref{fig-limbd} puts in evidence the full amount of
monochromatic limb
profile variations as a function of phase, for $\lambda =
500$~nm. As explained before, the limb darkening for our model of
$\zeta$~Gem is greater at minimum radius, and lower when the
shock-wave is crossing the photosphere.

The effect of phase on the limb profiles is not only restricted to
the amount of limb darkening, but is also reflected in the shape of
the profile itself. Consider the two profiles for $\phi \simeq 0.76$
and $\phi \simeq 0.91$. Despite the small difference in phase, they
cross each other at $\sin \alpha \simeq 0.75$, which indicates that
the gradient of $I_\nu$ with respect of $\sin \alpha$ is also a
function of the pulsational phase. This example highlights the intrinsic
limits, in the parameterization of the limb profiles, that result from
a single limb darkening parameter. 

A closer examination of figure~\ref{fig-limb} shows that, even when
the dynamic atmosphere is close to hydrostatic equilibrium ($\phi
\simeq 0.4$), static models are slightly less darkened for $\sin
\alpha \la 0.85$ and more darkened on the limb. Even in the most
favorable phases, the dynamic limb profiles cannot be obtained by
scaling suitable profiles computed with a static atmosphere. This is a
direct consequence of the similar difficulties, mentioned in
section~\ref{ssec-opc}, of fitting the hydrodynamic thermal structure
with a single hydrostatic model. Again, the different
functional shape between the static and dynamic profiles depends
strongly on the phase and the wavelength, and requires detailed
modeling. 

As expected, our limb profiles have larger darkening at shorter
wavelengths. This is mainly a consequence of the source function and
the optical depth dependence on wavelength. At 900~nm and in the
near-IR, different lines of sight along the stellar disk probe
regions of the stellar atmosphere where the source function is
similar. This results in a flat limb intensity profile, and thus low
limb darkening. At 500~nm the source function in the atmospheric
layers probed by the different directions change more rapidly towards
the limb, producing a larger limb darkening. This effect is even
larger when going to shorter wavelengths, in the UV (300--400~nm). 

This wavelength dependence of limb darkening is also affected by phase
dependent hydrodynamical effects. Figure~\ref{fig-ldcoeff} presents
limb profiles for $\lambda = 500$, 700 and 900~nm at two different
phases. As shown in the figure, the limb profiles derived for
dynamic atmospheres are in most cases very different from those
computed using static models, or from linear limb darkening curves
\citep{milne1921}.


\section{Conclusions}\label{sec-concl}

The increasing amount of available interferometric data requires
accurate stellar models to convert the raw
visibilities into meaningful angular diameters. This is
particularly true in the case of Cepheids, whose radial pulsations add
another degree of complexity in modeling their spectral emission. 
The Cepheid variability is being used as a means to accurately
determine their distance, and thus the scale of the entire universe;
at the same time there is a tendency to neglect
their pulsations when computing the properties of their
atmospheres, as they change during the Cepheid luminosity cycle.

We presented here a new approach that is used to derive time dependent
center-to-limb brightness distributions for classical Cepheids. The
adoption of second-order accurate hydrodynamical calculations in spherical
geometry, and the inclusions of a complete set of atomic and molecular
opacities, improve our limb darkening profiles over
previous computations. Conventional limb darkened interferometric
measurements are done by adopting a suitable stellar model having the
same spectral type and average \Teff{} of the Cepheid \citep[see
e.g.][]{nordgren1999}. In the past, however, none of the available
models, has been
specifically computed for a Cepheid, and the variations of \Teff{}
and \logg{} as the star pulsates are usually not taken into
account (the case of \logg{} is particularly dire, given the
difficulty of a precise measurement of this parameter, and \Teff{} may
suffer from the ambiguities in its definition).

These limitations are absent in our models, which can be computed
for individual Cepheids; the only requirements are an
accurate model for the pulsational piston \citep[see][]{buchler1990} and
the temporal variation of \Teff. Both conditions can be satisfied for
most of the nearby Cepheids, and constrained with available high
resolution spectral observations. 

The main differences between our dynamic models and conventional
static limb darkening profiles are analyzed in detail in the previous
section. However, it is important to remark once more that the amount
of limb darkening between static and dynamic models is significantly
different, and can be either larger or smaller according to the
pulsational phase. The actual shape of the limb profile is also a
function of phase, and cannot be reproduced by a single family of
static models, or a single limb darkening parameter. 

These results highlight the importance of hydrodynamic calculations in the
determination of center-to-limb brightness distribution for pulsating
Cepheids. In a future paper we will demonstrate
that the effects here described are already measurable with the current
interferometric technology, and affect the determination of Cepheid
angular diameters. This issue is becoming very important
when measuring the variations of such diameters as the star pulsates,
which is the first step to provide a reliable determination of its
distance through the geometric Baade-Wesselink method.


\acknowledgements
We wish to thank Bob Kurucz for valuable discussions and the
anonymous referee for comments which helped us to improve the paper. 
This work was partially supported by NSF grant AST9876734. M.K. is a
member of the Chandra Science Center, which is operated under contract
NAS8-39073, and is partially supported by NASA.

\clearpage


\clearpage


\figcaption[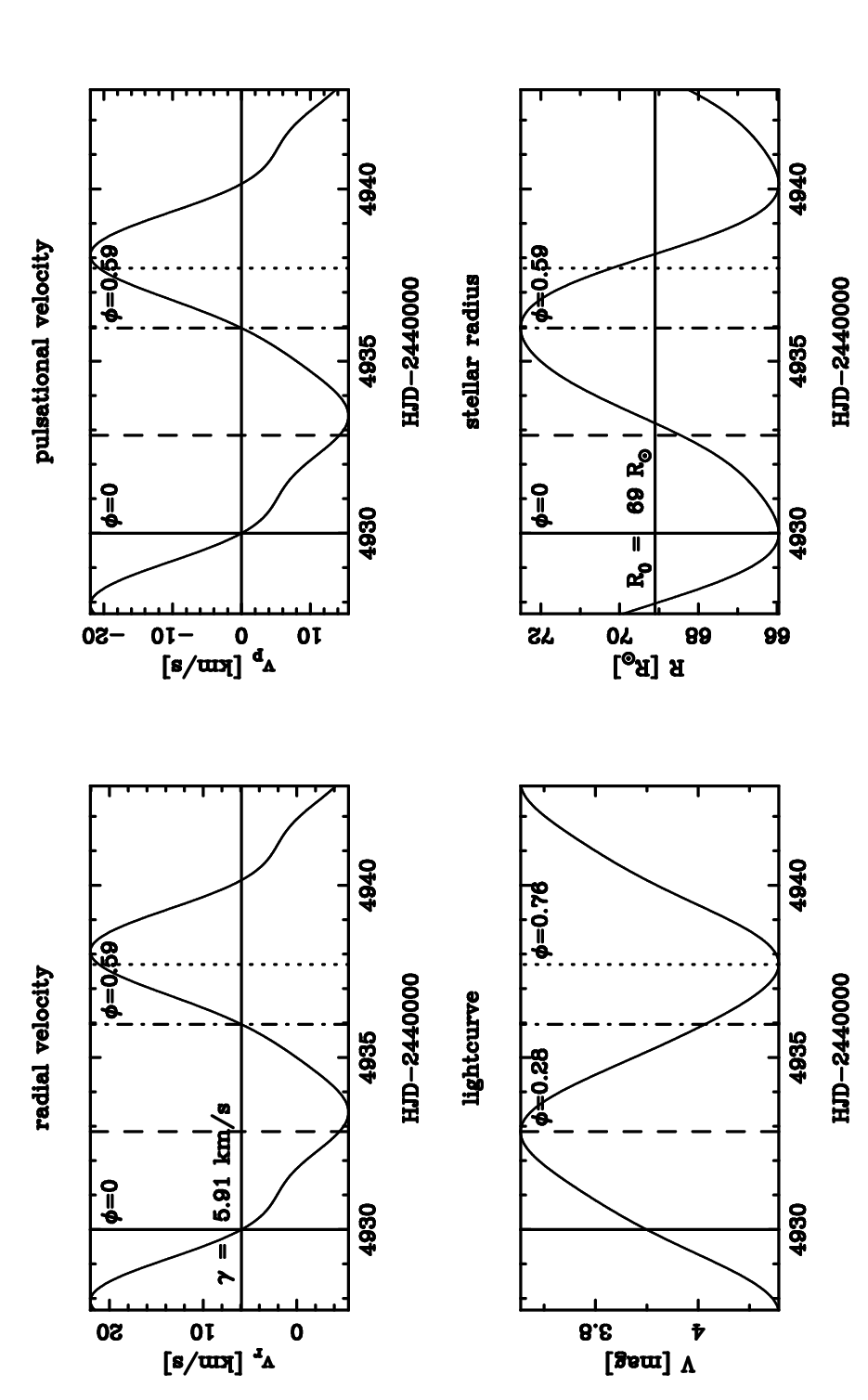]{Comparison between radial velocity, pulsational
velocity, lightcurve and radius for $\zeta$~Gem, based on
\citet{bersier1994a, bersier1994b}. The vertical solid line shows the
pulsational velocity zero phase; the lightcurve zero phase ($\phi
\simeq 0.28$) is indicated by the dashed line. Other important phases
are maximum radius ($\phi \simeq 0.59$ , dot-dashed line) and minimum
luminosity ($\phi \simeq 0.76$, dotted line).\label{fig-curves}}

\figcaption[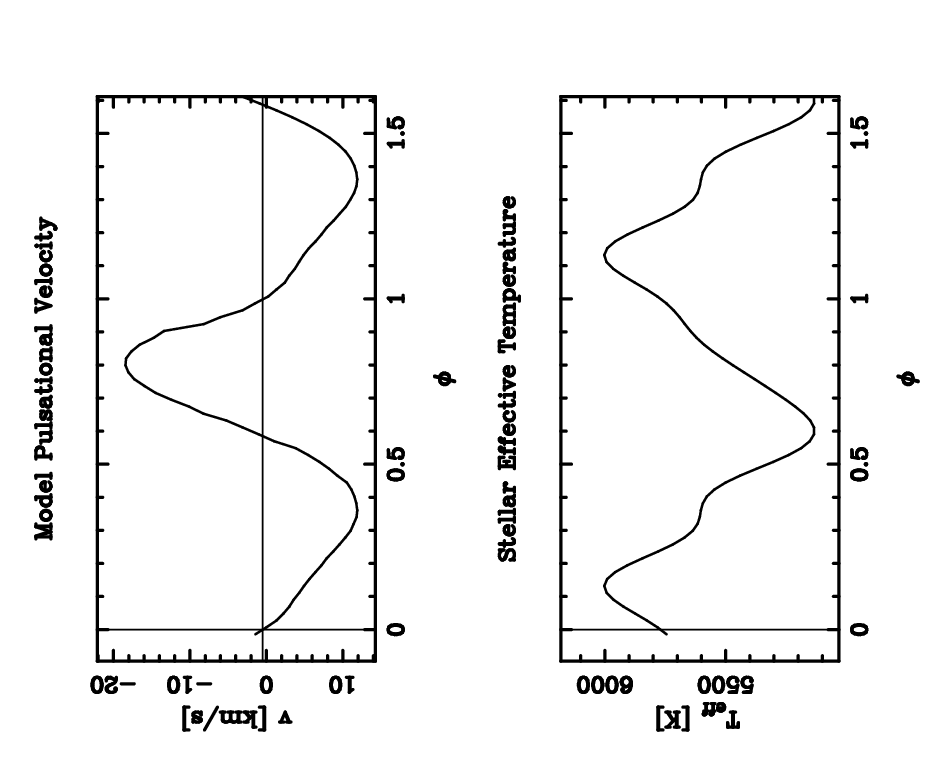]{Pulsational velocity, obtained by averaging the
velocities of the hydrodynamic model photospheric layers (top
panel). Bottom panel shows the effective temperature of $\zeta$~Gem,
derived as a function of phase by
\citet{krockenberger1997}.\label{fig-curvesk}}

\figcaption[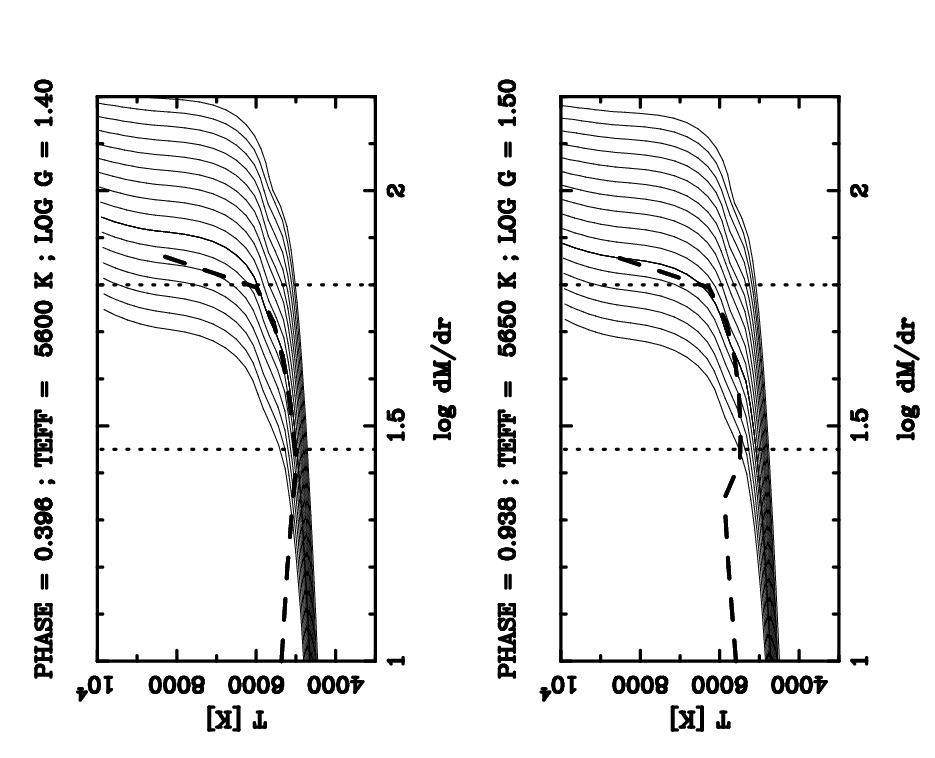]{Thermal structure of $\zeta$~Gem models. The
dashed lines are dynamic atmospheres computed at phase 0.4 
and at phase 0.94. Solid lines
are hydrostatic models having the same \Teff{} as the
hydrodynamic atmosphere, and \logg{} from 0.5 to 2.0 in step of
0.1. Note the temperature inversion in the dynamic models, occurring
for $\log \ud M / \ud r \simeq 1.45$, and the shock-wave crossing the
atmosphere at phase 0.94 ($\log \ud M / \ud r \la 1.3$). The
fitting region used to match dynamic atmospheres with static
models is marked by vertical dotted lines.\label{fig-A1}}

\figcaption[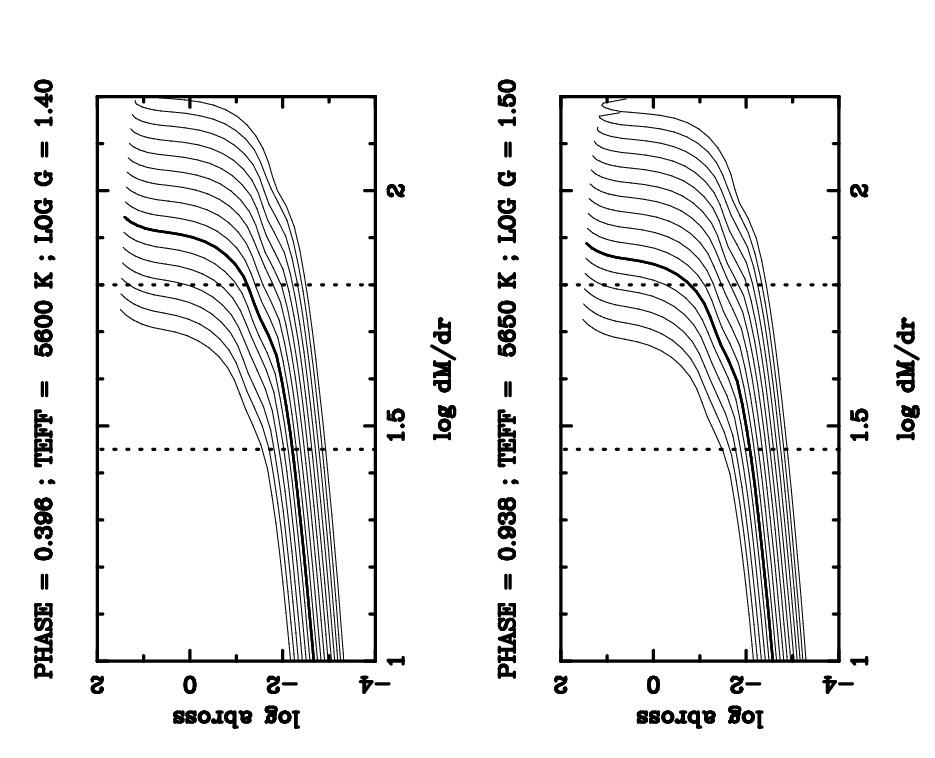]{Rosseland opacities for dynamic models at $\phi
\simeq 0.4$ and $\phi \simeq 0.94$ (top and bottom panels respectively).
The thick solid line is the Rosseland opacity of the static model
providing the best fit to the hydrodynamic atmosphere;
the thin lines are the opacities for other static models having the
same \Teff{} and different \logg. Given the uncertainty in the
determination of the best fit \logg, the resulting error in the 
Rosseland opacity can be visualized as the two curves bracketing
the best fit opacity. This is $\la \pm 0.3$ in the fitting region
(dotted vertical lines).\label{fig-A2}}

\figcaption[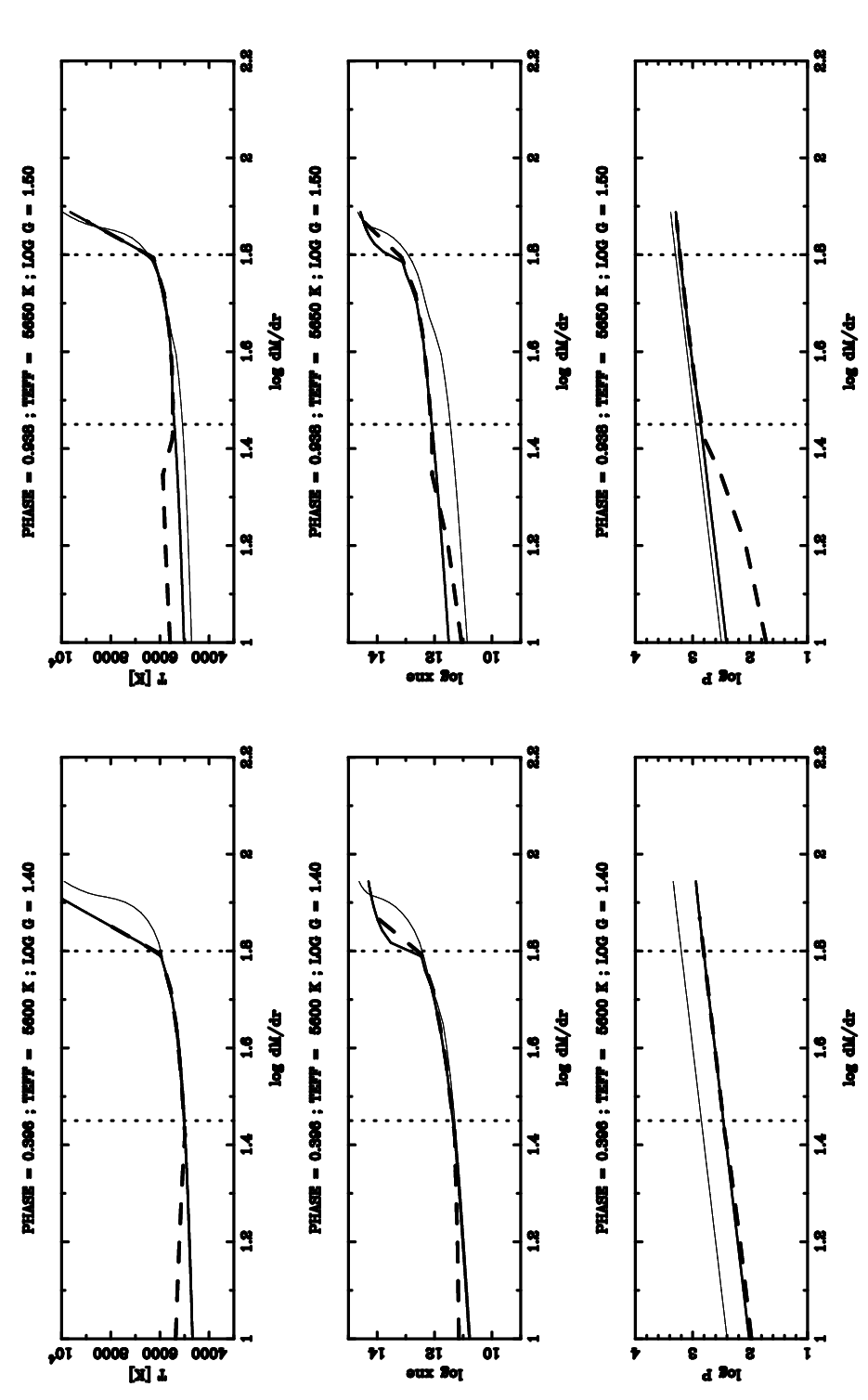]{Temperature, electron density and pressure of model
atmospheres at $\phi \simeq 0.4$ and $\phi \simeq 0.94$ (left and right
panels, respectively). The dashed lines are the original hydrodynamic
models, and the thin solid lines their best fit hydrostatic
atmospheres. The thick solid lines are the spliced ``hybrid'' models
obtained as explained in section~\ref{ssec-int}. Note that, at $\phi
\simeq 0.4$, the $T$ and $x_{\rm ne}$ distributions of static
and dynamic models are very similar, at least in the fitting
region (enclosed between the vertical dotted lines).
The pressure, on the contrary, is larger, due to the different
equation of state in the two models. The opposite is true at $\phi
\simeq 0.94$, because of the passage of the shock-wave, which excites
the atmosphere increasing $x_{\rm ne}$ with respect to the hydrostatic
equivalent.\label{fig-splice}}

\figcaption[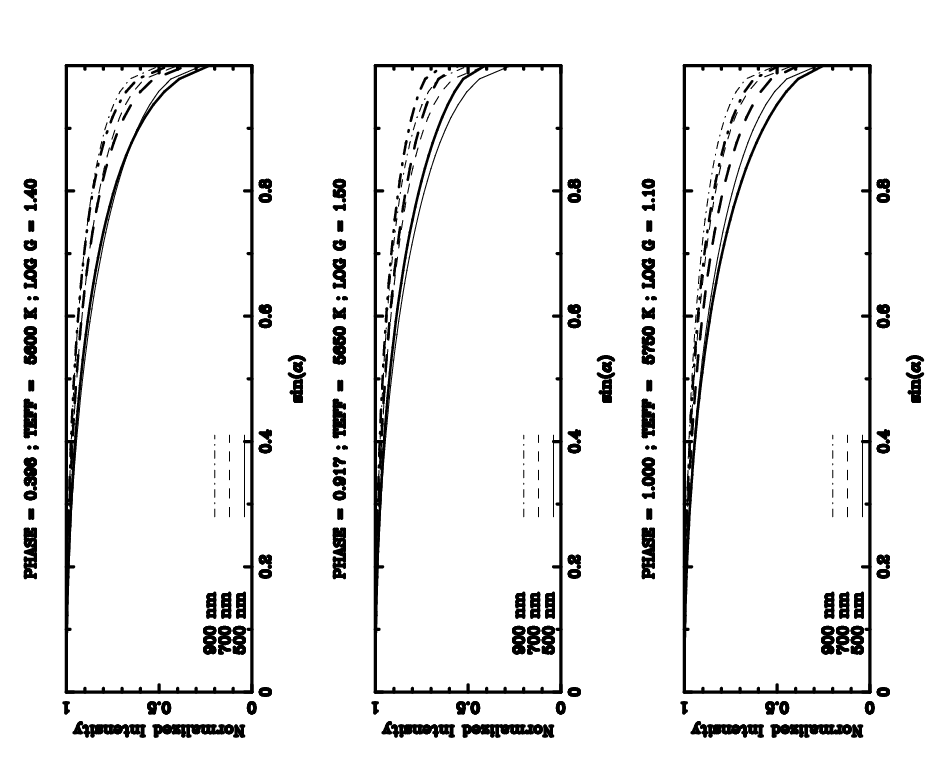]{Monochromatic limb profiles computed at 500, 700
and 900~nm for $\phi \simeq 0.4$, 0.92 and 1. Note that the low
spectral resolution of our intensity distributions is similar to the
spectral resolution of the actual NPOI interferometric
observations. Thick lines are profiles for the hydrodynamic model,
thin lines are computed for a hydrostatic atmosphere having the same
\Teff{} and \logg.\label{fig-limb}} 

\figcaption[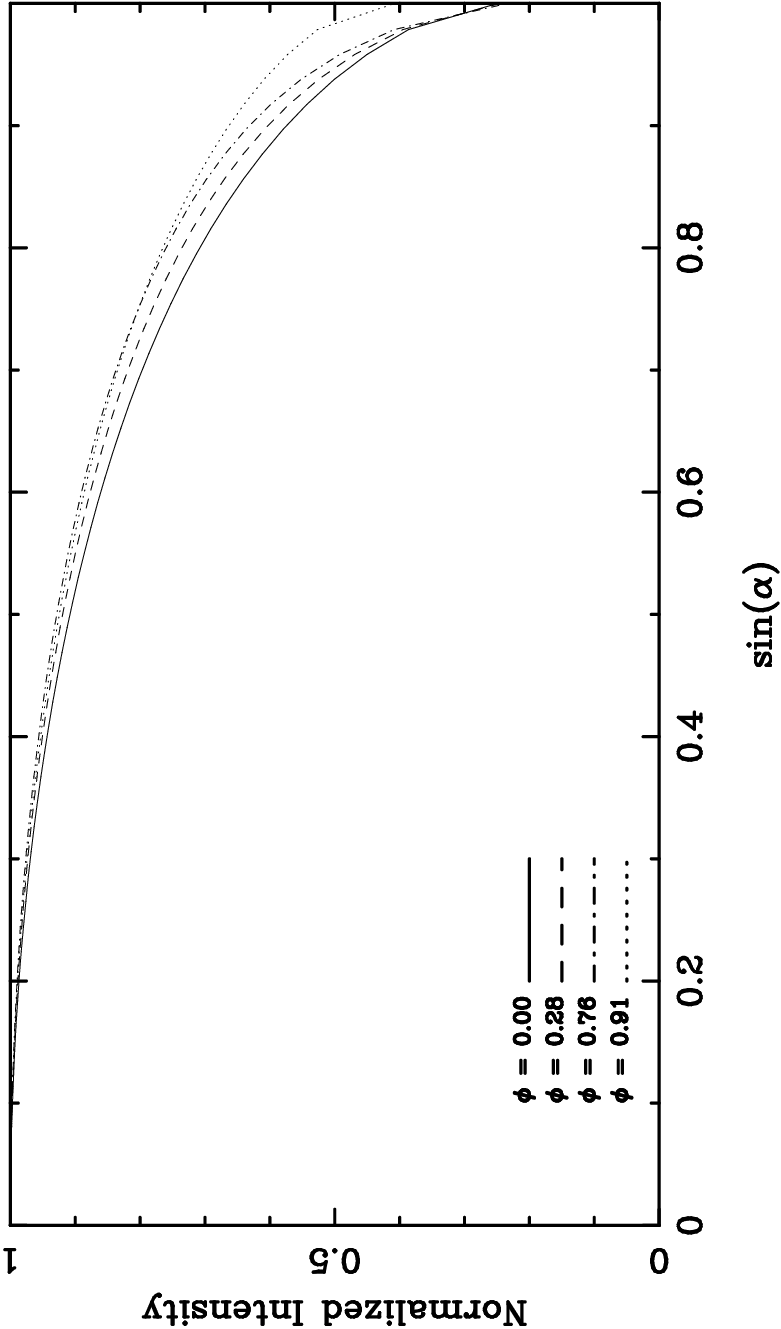]{Decreasing limb darkened profiles at phases 0
(minimum radius), 0.28 (maximum luminosity), 0.76 (minimum luminosity)
and 0.91 (passage of the shock-wave through the photosphere). The
profiles are computed for $\lambda = 500$~nm.\label{fig-limbd}}

\figcaption[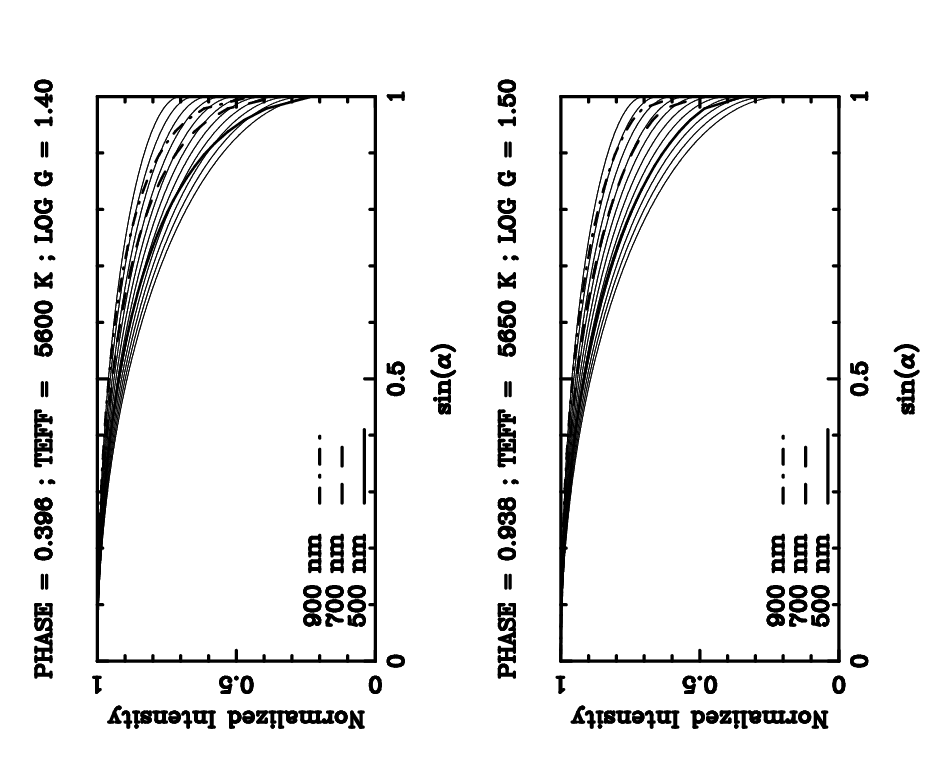]{Different levels of limb darkening at phase 0.4
and 0.94 for $\lambda = 500$, 700 and 900~\micron{} (thick
lines). Curves with linear limb darkening $I_\nu(\mu) = I_\nu(1) [1- u
(1 - \mu)]$, with $u$ from 0.3 to 0.80, are plotted
for reference (thin lines).\label{fig-ldcoeff}}

\clearpage


\epsscale{0.8}
\plotone{f1.eps}
\plotone{f2.eps}
\plotone{f3.eps}
\plotone{f4.eps}
\plotone{f5.eps}
\plotone{f6.eps}
\plotone{f7.eps}
\plotone{f8.eps}


\end{document}